\documentclass[11pt]{article}
\usepackage{moriond,epsfig}

\newcommand\as{\alpha_{\mathrm{S}}}

\def\slash#1{\ooalign{$\hfil/\hfil$\crcr$#1$}} 

\def\be{\begin{equation}}
\def\ee{\end{equation}}
\def\bea{\begin{eqnarray}}
\def\eea{\end{eqnarray}}

\begin{document}
\vspace*{4cm}
\title{HNNLO: a MC program for Higgs boson production at hadron colliders}

\author{ M. GRAZZINI }

\address{INFN, Sezione di Firenze, Via Sansone 1, I-50019 Sesto Fiorentino, Florence, Italy}

\maketitle\abstracts{We consider Higgs boson production through gluon--gluon fusion in hadron collisions. We present a numerical program that computes the cross section up to NNLO in QCD perturbation theory. 
The program allows the user to apply arbitrary cuts on the momenta of
the partons and of the photons or leptons that are
produced in the final state. We present selected numerical results at the Tevatron and the LHC.
}

\section{Introduction}

The Higgs boson is a fundamental ingredient in the Standard Model (SM) but
it has not been observed yet. Its search is currently being carried out at the Tevatron, and will be continued at the LHC.
Once the Higgs boson is found, the LHC will be able to study
its properties like couplings and decay widths.

At hadron colliders, the SM Higgs boson is mainly produced by
gluon-gluon fusion through a heavy-quark loop.
For such an important process, it is essential to have reliable theoretical
predictions for the cross section and the associated distributions.
At leading order (LO) in QCD perturbation theory,
the cross section is proportional
to $\as^2$, $\as$ being the QCD coupling. The QCD radiative corrections to the total cross section are known at the
next-to-leading order (NLO) 
\cite{Higgsnlo}
and at the next-to-next-to-leading order (NNLO) 
\cite{Higgsnnlo}.
The effects of a jet veto on the total cross section has been studied
up to NNLO \cite{Catani:2001cr}.
We recall that all the results at NNLO have been obtained by using 
the large-$M_t$ approximation, $M_t$ being the mass of the top quark.

These NNLO calculations can be supplemented with soft-gluon resummed calculations at next-to-next-to-leading logarithmic (NNLL) accuracy
either to improve the quantitative accuracy of the perturbative predictions
(as in the case of the total cross section \cite{Catani:2003zt,Moch:2005ky}) or to provide
reliable predictions in phase-space regions where fixed-order calculations
are known to fail \cite{qtresum}.

These fixed order and resummed calculations share a common feature: they
are fully inclusive over the produced final state (in particular, over
final-state QCD radiation).
Therefore they refer to situations
where the experimental cuts
are either ignored (as in the case of the total cross section) or taken into account
only in simplified cases (as in the case of the jet vetoed cross section).
The impact of higher-order corrections
may be strongly dependent on the details of the applied cuts and also the shape of various
distributions is typically affected by these details.

The first NNLO calculation that fully takes into account experimental cuts
was reported in Ref.~\cite{Hdiff}, considering the decay mode $H\to\gamma\gamma$.
In Ref.~\cite{Anastasiou:2007mz} the calculation is
extended to the decay mode $H\to WW\to l\nu l\nu$ (see also \cite{Anastasiou:2008ik}).

In Ref.~\cite{Catani:2007vq} we have
proposed a method to perform NNLO calculations and we have
applied it to perform an independent computation
of the Higgs production cross section.
The calculation is implemented
in a fully-exclusive parton level event generator.
This feature makes it particularly suitable for practical applications
to the computation of distributions in the form of bin histograms.
Our numerical program can be downloaded from \cite{hnnloweb}.
The decay modes that are currently implemented
are $H\to \gamma\gamma$ \cite{Catani:2007vq}, $H\to WW\to l\nu l\nu$ and $H\to ZZ\to 4$ leptons \cite{Grazzini:2008tf}.

In the following
we present a brief selection of results that can be obtained by our program.
We use the MRST2004 parton distributions \cite{Martin:2004ir},
with parton densities and $\as$ evaluated at each corresponding order
(i.e., we use $(n+1)$-loop $\as$ at N$^n$LO, with $n=0,1,2$). Unless stated otherwise,
the renormalization and factorization scales are fixed to the value 
$\mu_R=\mu_F=M_H$, where $M_H$ is the mass of the Higgs boson.

\section{Results at the LHC}

We consider the production of a Higgs boson with mass $M_H=165$ GeV
at the LHC ($pp$ collisions at $\sqrt{s}=14$ TeV)
in the decay mode $H\to WW\to l\nu l\nu$ \cite{Grazzini:2008tf}.
The NLO and NNLO inclusive $K$-factors are $K_{NLO}=1.84$ and $K_{NNLO}=2.21$, respectively.
We apply a set of selection cuts taken from the study of Ref.~\cite{Davatz:2004zg}. The charged leptons (with $|\eta|<2$) are classified according to their maximum ($p_{T1}$) and minimum ($p_{T2}$)
transverse momentum. Then $p_{T2}$ should be larger than 25 GeV, and
$p_{T1}$ should be between 35 and 50 GeV. The missing $p_T$ of the event is
required to be larger than $20$ GeV and the invariant mass of the charged
leptons must be smaller than $35$ GeV.
The azimuthal separation of the charged leptons in the
transverse plane ($\Delta\phi$) is smaller than $45^o$.
Finally, there should be no jet with $p_T^{\rm jet}$ larger than $30$ GeV.

The corresponding cross sections are reported in Table \ref{tab:wwsel}.
\begin{table}[htbp]
\begin{center}
\begin{tabular}{|c|c|c|c|}
\hline
$\sigma$ (fb)& LO & NLO & NNLO\\
\hline
\hline
$\mu_F=\mu_R=M_H/2$ & $17.36\pm 0.02$ & $18.11\pm 0.08$ & $15.70\pm 0.32$\\
\hline
$\mu_F=\mu_R=M_H$ & $14.39\pm 0.02$ & $17.07\pm 0.06$ & $15.99\pm 0.23$ \\
\hline
$\mu_F=\mu_R=2M_H$ & $12.00 \pm 0.02$ & $15.94\pm 0.05$ & $15.68\pm 0.20$\\
\hline
\end{tabular}
\end{center}
\caption{{\em Cross sections for $pp\to H+X\to WW+X\to l\nu l\nu+X$ at the LHC
when selection cuts are applied.}}
\label{tab:wwsel}
\end{table}

The cuts are quite hard, 
the efficiency being $8\%$ at NLO and $6\%$ at NNLO.
The scale dependence of the result is strongly reduced at NNLO,
being of the order of the error from the numerical integration.
The impact of higher order corrections is also
drastically changed. The $K$-factor is now 1.19 at NLO and
1.11 at NNLO.
As expected, the jet veto tends to stabilize the
perturbative expansion, and the NNLO cross section turns out to be smaller than the NLO one.
The study of Ref. \cite{Anastasiou:2008ik} shows that the efficiencies obtained at NNLO are in good
agreement with those predicted by the MC@NLO event generator \cite{MCatNLO}.


We now consider the production of
a Higgs boson with mass $M_H=200$ GeV \cite{Grazzini:2008tf}.
In this mass region the dominant decay mode is $H\to ZZ\to 4$ leptons,
providing a clean four lepton signature.
In the following we
consider the decay of the Higgs boson in two identical lepton pairs.
When no cuts are applied,
the NLO $K$-factor is $K_{NLO}=1.87$ whereas at NNLO we have $K_{NNLO}=2.26$.

We consider the following cuts \cite{CMStdr}.
The transverse momenta of the leptons, ordered from the largest ($p_{T1}$) to the smallest ($p_{T4}$) are required to fulfil $p_{T1}>30~{\rm GeV}$, $p_{T2}>25~{\rm GeV}$, $p_{T3}>15~{\rm GeV}$ and $p_{T4}>7~{\rm GeV}$.
The leptons should be central ($|\eta|< 2.5$)
and isolated: the total transverse energy $E_T$ in a cone of radius 0.2 around each lepton should fulfil $E_T< 0.05~p_T$.
For each possible $e^+e^-$ pair, the closest ($m_1$)
and next-to-closest ($m_2$) to $M_Z$ are found.
Then $m_1$ and $m_2$ are required to be $81$ GeV $< m_1 < 101$ GeV and $40$ GeV $< m_2 < 110$ GeV.
The corresponding cross sections are reported in Table \ref{tab:cuts}.
\begin{table}[htbp]
\begin{center}
\begin{tabular}{|c|c|c|c|}
\hline
$\sigma$ (fb)& LO & NLO & NNLO\\
\hline
\hline
$\mu_F=\mu_R=M_H/2$ & $1.541 \pm 0.002$ & $2.764\pm 0.005$ & $ 3.013\pm 0.023$\\
\hline
$\mu_F=\mu_R=M_H$ & $1.264\pm 0.001$ & $2.360\pm 0.003$ & $2.805\pm 0.015$\\
\hline
$\mu_F=\mu_R=2M_H$ & $1.047\pm 0.001$ & $2.044 \pm 0.003$ & $2.585\pm 0.010$\\
\hline
\end{tabular}
\end{center}
\caption{{\em Cross sections for $pp\to H+X\to ZZ+X\to e^+e^-e^+e^-+X$ at the LHC
when cuts are applied.}}
\label{tab:cuts}
\end{table}

Contrary to what happens in the $H\to WW\to l\nu l\nu$ decay mode,
the cuts are quite mild, the efficiency being $63\%$ at NLO and $62\%$ at NNLO.
The NLO and NNLO $K$-factors are $1.87$ and $2.22$, respectively.
Comparing with the inclusive case, we
conclude that these cuts do not change significantly
the impact of QCD radiative corrections.

\begin{figure}[htb]
\begin{center}
\begin{tabular}{c}
\epsfxsize=8truecm
\epsffile{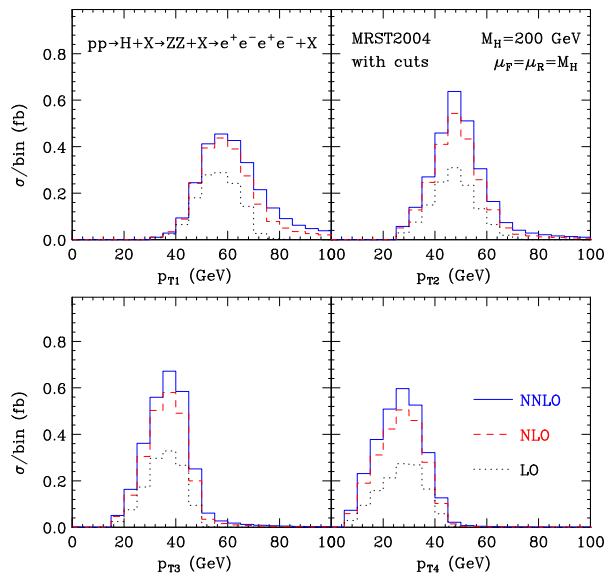}\\
\end{tabular}
\end{center}
\caption{\label{fig:ptlept}
{\em Tranverse momentum spectra of the final state leptons
for $pp\to H+X\to ZZ+X\to e^+e^-e^+e^-+X$,
ordered according to decreasing $p_T$,
at LO (dotted), NLO (dashed), NNLO (solid).}}
\end{figure}

In Fig.~\ref{fig:ptlept} we report the $p_T$ spectra of the charged leptons.
We note that at LO, without cuts, the $p_{T1}$ and $p_{T2}$ are kinematically bounded by $M_H/2$, whereas $p_{T3}< M_H/3$ and $p_{T4}<M_H/4$.
As is well known, in the presence of a kinematical boundary,
fixed order predictions may develop
perturbative instabilities \cite{Catani:1997xc}
at higher orders. In the present case,
the distributions smoothly reach the kinematical boundary
and no perturbative instability is observed beyond LO.

\section{Results at the Tevatron}

We now consider the production of a Higgs boson of mass $M_H=160$ GeV at the Tevatron ($p{\bar p}$ collisions at $\sqrt{s}=1.96$ TeV).
We consider the decay mode $H\to WW\to l\nu l^\prime \nu^\prime$
with $l,l^\prime=e,\mu$ and
use the cuts from \cite{Abulencia:2006aj}.
The inclusive $K$-factors at NLO and NNLO are $K_{NLO}=2.42$
and $K_{NNLO}=3.31$, respectively.
The trigger requires either a central electron with
$|\eta|<1.1$ and $E_T>18$ GeV, or a forward electron with $1.2<|\eta|<2$ with $E_T>20$ GeV and
$\slash E_T>15$ GeV, or a central muon with $|\eta|<1$ and $p_T>18$ GeV.
The leptons should be isolated: the energy in a cone of radius $R=0.4$ around each lepton should fulfill $E<0.1~p_T$.
The selection cuts require $p_{T1}>30~{\rm GeV}$, $p_{T2}>25~{\rm GeV}$ and
$\slash E_T>40$ GeV. If $\slash E_T<50$ GeV the azimuthal separation $\Delta\phi(\slash E_T,p)$ should be larger than $20^o$ for each lepton or jet.
The invariant mass of the charged leptons should be between 16 and 75 GeV.
The scalar sum of the $p_T$ of the leptons and of $\slash E_T$ should be smaller than $M_H$.
Finally, jets with $E_T>15$ GeV and $|\eta|<2.5$ are counted. We require either no such jet,
or one of such jets with $E_T$ smaller than 55 GeV, or two of such jets with $E_T$ smaller than
40 GeV. Since the small $\Delta\phi$ region is the most important,
we further require $\Delta\phi< 80^o$.

With these cuts the LO, NLO and NNLO cross sections are $\sigma_{LO}=1.571\pm 0.003$, $\sigma_{NLO}=3.16\pm 0.01$, $\sigma_{NNLO}=2.78\pm 0.17$ fb.
The corresponding $K$-factors are $K_{NLO}=2.01$ and $K_{NNLO}=1.77$, and exibit a strong reduction with respect to the inclusive case. As for the LHC, the selection cuts drastically change the impact of radiative corrections.
Also the efficiencies are reduced: comparing
the accepted and inclusive cross sections we obtain:
$\epsilon_{LO}=33\%$, $\epsilon_{NLO}=27\%$ and $\epsilon_{NNLO}=18\%$.
These results suggest the existence of
large theoretical uncertainties that need to be further
investigated.

\end{document}